# Modeling Interconnect Variability Using Efficient Parametric Model Order Reduction

Peng Li, †Frank Liu, ‡Xin Li, ‡Lawrence T. Pileggi and †Sani R. Nassif

Department of EE, Texas A&M University, USA, †IBM Austin Research Laboratories, USA,
‡Department of ECE, Carnegie Mellon University, USA

## Abstract

Assessing IC manufacturing process fluctuations and their impacts on IC interconnect performance has become unavoidable for modern DSM designs. However, the construction of parametric interconnect models is often hampered by the rapid increase in computational cost and model complexity. In this paper we present an efficient yet accurate parametric model order reduction algorithm for addressing the variability of IC interconnect performance. The efficiency of the approach lies in a novel combination of low-rank matrix approximation and multi-parameter moment matching. The complexity of the proposed parametric model order reduction is as low as that of a standard Krylov subspace method when applied to a nominal system. Under the projection-based framework, our algorithm also preserves the *passivity* of the resulting parametric models.

## 1. Introduction

The every-increasing variability of modern IC manufacturing process has introduced significant variations in circuit performances. Accordingly, assessment of these performance variations has become an integral part of analysis and optimization of modern VLSI designs. In the past decade, efficiency in IC interconnects analysis has been facilitated by powerful model order reduction techniques [1]-[4]. Two classes of model order reduction techniques have gained popularity: moment-matching based method (e.g. AWE [1] and Krylov subspace methods [2]-[4]) and control-theoretic approaches (e.g. truncated balanced realization (TBR) [5][8][11]). The first class of methods are very attractive in terms of computational cost while the methods that fall into the second class tend to be more accurate, but suffer from a dramatic increase in computational cost. The high cost associated with the latter often preclude these methods from being directly applied to large practical problems.

Variational modeling of interconnects can be classified similarly: moment-matching (Krylov subspace) based [6] vs. TBR-based [7][12][13]. In [6], Liu et. al. proposed to construct variational models by fitting the projection vectors over the samples taken in the variational parameter space. In [7], Heydari et. al. performed TBR analysis on perturbed systems. Additionally, a linear fractional transform (LFT) formulation based TBR-like approach was proposed in [12][13]. Almost the same comparison can be drawn for these variational methods: moment matching based approaches are computationally more efficient while TBR methods tend to be more accurate but with a significantly higher cost.

Nevertheless, in comparison to modeling techniques employed for the nominal system, the inclusion of variational effects into the model order reduction implies a significant increase in computational cost and/or model complexity. This can be true even for the more efficient moment matching methods. Hence, in this paper we focus on developing more efficient Krylov subspace methods.

In this paper, we first show that in principle, variational modeling of interconnects can be cast as a general multi-parameter Krylov-based moment matching problem [10]. We then point out that for the particular problem of modeling process variations this is not an entirely suitable choice due to its limitation in model complexity. As an improvement, we show that exploiting multi-point expansion under the same Krylov framework can lead to more compact interconnect models but with an increase in computational cost. Furthermore, we propose a novel combination of low-rank matrix approximation and multi-parameter moment matching scheme and show that we can maintain the same low cost of standard Krylov methods while significantly improve model complexity. The new approach allows us to achieve a higher moment-matching order for multi-parameter moments than what was possible before, thereby significantly improving the model accuracy. One further advantage of the approach is that the *passivity* of reduced parametric models can be easily guaranteed. We demonstrate the proposed technique on several variational interconnect modeling examples.

## 2. Prior Work

An interconnect network can be described by the following MNA (modified nodal analysis) formulation:

$$C\dot{x} = -Gx + Bu, \quad y = L^T x \qquad (1)$$

where $G$, $C$, $B$ and $L$ are system matrices, $u \in R^m$ and $y \in R^m$ denote port voltages and currents, $x \in R^n$ is the vector consisting of nodal voltages and branch currents for voltage sources and inductors. The PRIMA algorithm of [4] computes an orthonormal basis $V \in R^{n \times q}$ for a Krylov subspace of $A \equiv -G^{-1}C$, which is spanned by several block moments of the system transfer function. The reduced order model of (1) is obtained by computing the following matrices of smaller dimensions:

$$\tilde{G} = V^T G V,\ \tilde{C} = V^T C V,\ \tilde{B} = V^T B,\ \tilde{L} = V^T L\ . \qquad (2)$$

To model the variational effects of interconnects, [6] uses the first order expansions to express the system conductance and capacitance matrices:

$$G(p_1, p_2) = G_0 + p_1 G_1 + p_2 G_2,\ C(p_1, p_2) = C_0 + p_1 C_1 + p_2 C_2 \quad (3)$$

where $G_0$ and $C_0$ are conductance and capacitance matrices under the nominal conditions, $p_1$ and $p_2$ are parameters used to model variational sources such as the metal line width or thickness variations. For convenience, we refer to $p_1$ and $p_2$ as variational parameters, and $G_1$, $C_1$, $G_2$, $C_2$ the sensitivity matrices w.r.t variational parameters. When the circuit parameters vary due to process fluctuations, the projection matrix used in PRIMA also undergoes small variations. The projection matrix is therefore expanded using a Taylor series [6]:

$$V(p_1, p_2) = V_0 + V_{11}p_1 + V_{21}p_1^2 + V_{12}p_2 + V_{22}p_2^2 \quad (4)$$

To determine the coefficient matrices in the above equation, samples are taken in the variational parameter space and the PRIMA algorithm is applied to the resulting perturbed systems. Based on these sampled projection matrices, coefficient matrices in (4) are determined by solving a set of linear equations. Inserting (4) into (2), a variational reduced order model parametrizable in $p_i's$ is obtained.

# 3. Multi-Parameter Moment Matching

In this section, we first point out that parametric reduced order modeling can be more generally cast as a *multi-parameter* Krylov projection based moment matching problem. In what follows, we show the benefit of exploiting multi-point expansions under this context.

## 3.1 Single-point expansion

A Krylov projection based multi-parameter moment matching approach was adopted in [10] to generate parameterized interconnect performance models. We shall refer to the approach in [10] as a *single-point* expansion method since the system transfer function is expanded at only one point in the joint space of *s* and $p_i's$. To see how multi-parameter moment matching can be applied, let us consider a more general case of (3) where conductance and capacitance matrices depend on $n_p$ parameters

$$\begin{aligned} G(p_1, p_2, \cdots, p_{n_p}) &= G_0 + p_1G_1 + p_2G_2 + \cdots + p_{n_p}G_{n_p} \\ C(p_1, p_2, \cdots, p_{n_p}) &= C_0 + p_1C_1 + p_2C_2 + \cdots + p_{n_p}C_{n_p} \end{aligned} \quad (5)$$

We use the notation $\{G_0, C_0, G_1, \cdots, G_{n_p}, C_1, \cdots, C_{n_p}, B, L\}_{n_p}$ to specify a $n_p$-parameter dynamic system as in (5). The transfer function of the system can be written as

$$\begin{aligned} & X(s, p_1, p_2, \cdots, p_{n_p}) \\ &= \left[G(p_1, p_2, \cdots, p_{n_p}) + sC(p_1, p_2, \cdots, p_{n_p})\right]^{-1} B \\ &= \left[I + sG_0^{-1}C_0 + \sum_{i=1}^{n_p} p_i G_0^{-1} G_i + \sum_{i=1}^{n_p} sp_i G_0^{-1} C_i\right]^{-1} G_0^{-1} B \end{aligned} \quad (6)$$

(6) can be readily expanded into a power series in several variables $s, p_1, p_2, \cdots, p_{n_p}$

$$\begin{aligned} & X(s, p_1, p_2, \cdots, p_{n_p}) \\ &= \sum_{k=0}^{\infty} \left[- sG_0^{-1}C_0 - \sum_{i=1}^{n_p} p_i G_0^{-1} G_i - \sum_{i=1}^{n_p} sp_i G_0^{-1} C_i\right]^k G_0^{-1} B \\ &= \sum_{k=0}^{\infty} \sum_{k_s=0}^{k} \sum_{k_1=0}^{k-k_s} \sum_{k_2=0}^{k-k_s-k_1} \cdots \sum_{k_{n_p}=0}^{k-k_s-k_1-\cdots k_{n_q-1}} M_{k,k_s,k_1,\cdots,k_{n_p}} s^{k_s} p_1^{k_1} \cdots p_{n_p}^{k_{n_p}} \end{aligned} \quad (7)$$

In (7) $M_{k,k_s,k_1,\cdots,k_{n_p}} (k_s + k_1 + \cdots k_{n_p} = k)$ is a *k*-th order multi-parameter moment corresponding to the coefficient of the term $s^{k_s} p_1^{k_1} p_2^{k_2} \cdots p_{n_p}^{k_{n_p}}$. Consider an orthonormal basis $V \in R^{n \times q} (q < n)$ of the subspace spanned by the multi-parameter moments with an order equal to or less than *k*

$$colspan(V) = span\{M_{0,0,\cdots 0}, M_{1,1,\cdots 0}, \cdots, M_{k,k_s,k_1,\cdots k_{n_p}}, \cdots\} \quad (8)$$

We can employ *V* as a projection matrix to compute a reduced model; i.e., each system matrix of the full model including sensitivity matrices is reduced using matrix transformations of (2). For instance, the sensitivity of $G_{p_k}$ is reduced to $\tilde{G}_{p_k} = V^T G_{p_k} V$. It can be shown that the resulting reduced model matches up to the *k*-th order multi-parameter moments of the original transfer function.

## 3.2 Inefficiency of single-point expansion

Although all the parameters are treated identically in the above multi-parameter moment matching scheme, there does exist an *asymmetry* between these parameters that arise practically in our *manufacturing-variation-aware* interconnect modeling problem. While variational parameters fluctuate in a small range around the nominal values due to process variations, the "*variation*" in the frequency variable *s* has to be considered over a wider range in order to capture the complete system frequency response.

It is not surprising that the reduced order model size of the above approach grows rapidly with the number of parameter and the orders of moment matching. With more careful analysis, it is quite interesting to see that the model size actually depends not so much on the moment matching order of individual parameters but the highest moment matching order with respect to any parameter. This is because a high moment matching order of any parameter will introduce a large number of cross terms in the multi-parameter expansion, thus lead to a large reduced model size. In other words, for interconnect variational modeling, choosing a low moment matching order for variational parameters (due to the small range of variation) does not significantly reduce the model size if the order for variable *s* remains high. As a result, the single-point expansion produces an *expensive* reduced model even when the circuit is only perturbed within a small range about the nominal condition.

## 3.3 Multi-point expansion

Let us consider the multi-point expansion in Fig. 1, where multiple samples are taken within the specified ranges in the variational parameter space. Let us assume that $n_s$ samples are taken and for each perturbed system $P_i = [p_1 = p_{i,1}, p_2 = p_{i,2}, \cdots, p_{n_p} = p_{i,n_p}]^T$, a projection matrix $V_i$ is computed to match the *k* moments of *s* using any standard Krylov method. Finally, a parametric model can be produced by forming an orthonormal basis *V* of the combined projection vectors $V = colspan\{V_1, V_2, \cdots, V_{n_s}\}$ and employing $V$ as the final projection matrix. The resulting model preserves the first $k$ moments of $s$ at the above $n_s$ sample points. In some sense, the reduced model approximates the full model at the sample points using a standard model order reduction technique, and then interpolates implicitly between these samples in the variational parameter space.

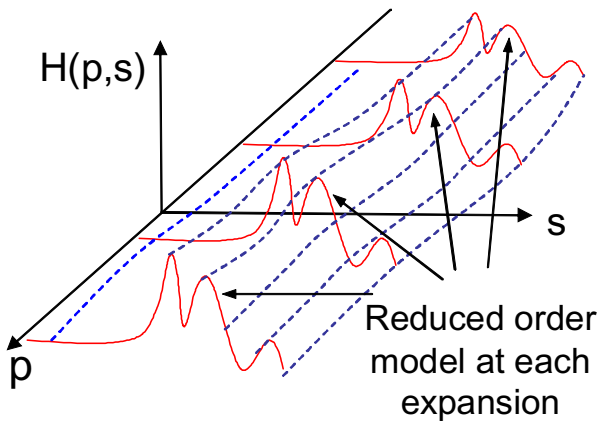

**Fig 1. Multi-point expansion in the process parameter space.**

In addition to other benefits of a multi-point method, it can be used to reduce the model size. To see this, let us assume that we want to match multi-parameter transfer function moments corresponding to the following terms: $s^0 p_i^0, s^1 p_i^0, s^2 p_i^0, \cdots, s^k p_i^0, p_i, sp_i, s^2 p_i, \cdots, s^k p_i$, where the expansion order for some variational parameter $p_i$ is the 1st order. To do so, using a single-point method will lead to a reduced order model of size $(k^2 + k + 1)m$, where $m$ is the number of ports. In a quite equivalent way, one can employ a multi-point method to match the first $k+1$ moments of $s$ at two distinct samples of $p_i$. This gives a much smaller reduced model of size $2(k+1)m$. A similar situation is encountered under the context of nonlinear system model order reduction [9].

The aforementioned multi-point method bears resemblance of the approach of [6]. The major difference is that in [6] interpolation between the sample points is done by direct fitting while in this approach it is accomplished implicitly via projection. Sometimes it is observed that the projection matrix is sensitive w.r.t variational parameters thus making a direct fitting less robust. Under these cases, multi-point expansion might be a more robust choice.

## 4. Low-Rank Approximation Based Multi-Parameter Moment Matching

### 4.1 The proposed approach

Although sampling in variational parameter space can lead to an improvement in model complexity, the computational cost and/or model size can still be high when several variational parameters are considered simultaneously. For instance, if we sample a four-dimensional parameter space by taking three samples per axis, a total of 81 sample points will result that requires the same number of matrix factorizations. To see the interactions between different parameters, we rewrite the transfer function expansion in (7)

$$X(s, p_1, p_2, \cdots, p_{n_p}) = \left( I + \sum_{k=1}^{\infty} \sum_{i=1}^{n_p} \sum_{j=1}^{n_p} \bigcup_{l=1}^{k} \left( \sum_{a_l, g_l, c_l,} s^{a_l + c_l} p_i^{g_l} p_j^{c_l} A_0^{a_l} \left(-G_0^{-1} G_i\right)^{g_l} \left(-G_0^{-1} C_j\right)^{c_l} \right) \right) R_0 \quad (9)$$

where $A_0 = -G_0^{-1} C_0$, $R_0 = G_0^{-1} B$, $a_l, g_l, c_l \in \{0,1\}$ and $a_l + g_l + c_l = 1$. We refer to each $G_0^{-1} G_i$, $G_0^{-1} C_i$, $1 \le i \le n_p$ as a *generalized sensitivity matrix* with respect to $G_0$.

Consider one possible term in the summation of (9)

$$F = s^{t+q} p_i A_0^t \left(-G_0^{-1} G_i\right) A_0^q R_0 \quad (10)$$

To match the moment expression in (10) more efficiently, we first seek the optimal 2-norm rank-$k_{G_i}$ ($k_{G_i} << n$) approximation of the generalized sensitivity matrix $G_0^{-1} G_i$ using SVD (singular value decomposition) [16]

$$-G_0^{-1} G_i \approx \sum_{j=1}^{k_{G_i}} \sigma_j u_j v_j^T = \hat{U}_{G_i} \cdot \hat{V}_{G_i}^T, \quad (11)$$
$$\hat{U}_{G_i} = [\sigma_1 u_1, \cdots, \sigma_{k_{G_i}} u_{k_{G_i}}], \hat{V}_{G_i} = [v_1, \cdots, v_{k_{G_i}}]$$

where $\sigma_j, u_j, v_j, 1 \le j \le k_{G_i}$ are the $k_{G_i}$ largest singular values and the corresponding left and right singular value vectors, respectively. Substituting (11) into (10) yields

$$\begin{aligned} F &\cong s^{t+q} p_i A_0^t \hat{U}_{G_i} \hat{V}_{G_i}^T A_0^q R_0 \\ &= s^{t+q} p_i A_0^t \hat{U}_{G_i} \hat{V}_{G_i}^T \overbrace{(-G_0^{-1} C_0)(-G_0^{-1} C_0) \cdots (-G_0^{-1} C_0)}^{q \text{ terms}} R_0 \\ &= s^{t+q} p_i A_0^t \hat{U}_{G_i} \hat{V}_{G_i}^T (-G_0^{-1})(-C_0 G_0^{-1})^{q-1} C_0 R_0 \end{aligned} \quad (12)$$

Now defining $\tilde{V}_{G_i} = -G_0^{-T} \hat{V}_{G_i}$ and $A_{0T} = -G_0^{-T} C_0^T$, we have

$$\begin{aligned} F &\cong s^{t+q} p_i A_0^t \hat{U}_{G_i} (-\hat{V}_{G_i}^T G_0^{-1})(-C_0 G_0^{-1})^{q-1} C_0 R_0 \\ &= s^{t+q} p_i A_0^t \hat{U}_{G_i} \left[(-G_0^{-T} C_0^T)^{q-1} (-G_0^{-T} \hat{V}_{G_i})\right]^T C_0 R_0 \\ &= s^{t+q} p_i A_0^t \hat{U}_{G_i} \left[(A_{0T})^{q-1} \tilde{V}_{G_i})\right]^T C_0 R_0 \end{aligned} \quad (13)$$

There are two important implications due to the approximation in (13). First, it suggests that there are two useful Krylov subspaces, one w.r.t $A_0$ and the other w.r.t $A_{0T}$ that can be used for matching the corresponding moment:

$$\begin{aligned} Kr(A_0, \hat{U}_{G_i}, t+1) &= colspan\{\hat{U}_{G_i}, A_0 \hat{U}_{G_i}, \cdots, A_0^t \hat{U}_{G_i}\} \\ Kr(A_{0T}, \tilde{V}_{G_i}, q) &= colspan\{\tilde{V}_{G_i}, A_{0T} \tilde{V}_{G_i}, \cdots A_{0T}^{q-1} \tilde{V}_{G_i}\} \end{aligned} \quad (14)$$

Secondly, the use of low-rank matrix approximation for generalized sensitivity matrices has decoupled the Krylov subspace construction w.r.t various parameters, i.e. these subspaces can be computed independently from each other and combined at a later time to build the overall projection matrix. Although we have only shown this decoupling between frequency variable $s$ and parameter $p_i$ in the simple case of (10) due to the space limitation, the same property remains even for an arbitrary combination of multiple parameters. This comes from the fact that we include the Krylov subspaces associated with all the low-rank approximation vectors in the projection. The immediate benefit of this result is that the size of the reduce order model can be significantly reduced as the cross terms of multi-parameter moments are no longer an issue.

It is worth mentioning that a similar reduction in model size can be achieved if low-rank approximations are applied to sensitivity matrices instead of generalized sensitivity matrices. However, this choice will incur a larger error in approximating the true transfer function. We have observed that approximating the generalized sensitivity matrices work much better in practice due to their stronger connection to moments, as evident in (9). We outline the proposed low-rank approximation based single-point multi-parameter moment matching scheme in algorithm 1 (Fig. 2). We summarize the moment matching property of algorithm 1 in theorem 1. Due to the space limitation, we omit the proof of the theorem.

**Theorem 1** (moment matching property of algorithm 1):

*For the approximate parametric system* $\{G_0, C_0, \tilde{G}_1, \cdots, \tilde{G}_{n_p}, \tilde{C}_1, \cdots, \tilde{C}_{n_p}, B, L\}_{n_p}$, *where* $\tilde{G}_i = -G_0 \hat{U}_{G_i} \hat{V}_{G_i}^T$, $\tilde{C}_i = -G_0 \hat{U}_{C_i} \hat{V}_{C_i}^T$, $1 \le i \le n_p$, *the reduced order model using the projection matrix computed in algorithm 1 preserves the transfer function moments with respect to all parameters up to order k.*

As in Theorem 1, the projection *V* computed in Algorithm 1 corresponds to a nearby system that is based on the low-rank approximation of generalized sensitivity matrices. In other words, we compute an approximate projection matrix for the original parametric system and employ it for parametric model order reduction. It should be noted that the same moment matching property of theorem 1 can be achieved by neglecting the Krylov subspaces with respect to $A_{0T}$, namely $V_{G_i,2}$ and $V_{C_i,2}$ in step 2.2 of the algorithm, but adding $\hat{V}_{G_i}$ and $\hat{V}_{C_i}$ to the projection matrix. This simplification can reduce the model size approximately by a factor of two. Nevertheless, once the projection matrix is computed we will apply matrix reduction on the sensitivity matrices but *not* their low-rank approximations as in step 4 of the algorithm. Therefore, incorporating the useful Krylov subspaces of $A_{0T}$ improves the accuracy of the model order reduction. The congruence transforms employed in step 4 of the algorithm also implies that the *passivity* of the reduced model will be guaranteed if the original parametric model is passive.

## 4.2 Computational cost and model complexity

One appealing property of Algorithm 1 is that the dominant cost of the parametric model order reduction is the same as that of a Krylov subspace method when applied to a deterministic system. The dominant cost is again one-time factorization of a sparse matrix, $G_0$.

Although the generalized sensitivity matrices are dense, the computation of a few dominant singular values/vectors for them (step 1 of algorithm 1) can be efficiently facilitated by the matrix-implicit nature of iterative sparse SVD techniques [14][15]. For instance, a low-rank approximation of $-G_0^{-1} G_i^T$ can be efficiently done using a few subspace iterations wherein the dense generalized sensitivity matrix is not explicitly required but only its matrix-vector products. The latter can be provided efficiently by reusing the same factorization of $G_0$. Furthermore, the computation of Krylov subspaces with respect to $A_{0T}$ can be achieved by the same sharing. Notice that if the LU factorization of $G_0$ is $G_0 = L_g U_g$, then $G_0^T = U_g^T L_g^T$. The matrix-vector product $y = A_{0T} \cdot x$ can be achieved by solving $-G_0^T y = C_0^T x$ or $-U_g^T L_g^T y = C_0^T x$. Therefore, the cost of the algorithm is linear in both the moment matching order *k* and the number of variational parameters $n_p$, and almost linear in the number of circuit nodes.

For a *k*-th order multi-parameter moment matching, the size of reduced order model is $O(4k_{svd} n_p + m)k$, where $k_{svd}$ is the rank of the SVD approximation, and *m* the number of ports (using the simplification pointed out in section 4.1

**Algorithm 1:**

**1.** *Compute a low-rank approximation for each generalized relative sensitivity matrix:*

$-G_0^{-1} G_i \approx \hat{U}_{G_i} \hat{V}_{G_i}^T$, $-G_0^{-1} C_i \approx \hat{U}_{C_i} \hat{V}_{C_i}^T$, $1 \le i \le n_p$

**2.** *Compute an orthonormal basis of the following Krylov subspaces*:

**2.1.** $V_0 = colspan\ \{R_0, A_0 R_0, \cdots, A_0^k R_0\}$
*where* $A_0 = -G_0^{-1} C_0, R_0 = G_0^{-1} B$

**2.2.** *For each* $i$, $1 \le i \le n_p$,

$\tilde{V}_{G_i} = -G_0^{-T} \hat{V}_{G_i}$, $\tilde{V}_{C_i} = -G_0^{-T} \hat{V}_{C_i}$

$V_{G_i,1} = colspan\ \{\hat{U}_{G_i}, A_0 \hat{U}_{G_i}, \cdots, A_0^k \hat{U}_{G_i}\}$

$V_{G_i,2} = colspan\ \{\tilde{V}_{G_i}, A_{0T} \tilde{V}_{G_i}, \cdots, A_{0T}^{k-1} \tilde{V}_{G_i}\}$

$V_{C_i,1} = colspan\ \{\hat{U}_{C_i}, A_0 \hat{U}_{C_i}, \cdots, A_0^{k-1} \hat{U}_{C_i}\}$

$V_{C_i,2} = colspan\ \{\tilde{V}_{C_i}, A_{0T} \tilde{V}_{C_i}, \cdots, A_{0T}^{k-2} \tilde{V}_{C_i}\}$

*where* $A_{0T} = -G_0^{-T} C_0^T$

**3.** *Compute an othonormal basis V for the combined subspaces:*

$V = colspan\ \{V_0, V_{G_1,1}, V_{G_1,2}, V_{C_1,1}, V_{C_1,2}, \cdots, V_{G_{n_p},1}, V_{G_{n_p},2}, V_{C_{n_p},1}, V_{C_{n_p},2}\}$

**4.** *Construct the final reduced order model:*

$\tilde{G}_0 = V^T G_0 V$, $\tilde{C}_0 = V^T C_0 V$, $\tilde{B} = V^T B$, $\tilde{L} = V^T L$,

$\tilde{G}_i = V^T G_i V$, $\tilde{C}_i = V^T C_i V$, *for* $1 \le i \le n_p$

**Fig 2. Low-rank approximation based single-point multi-parameter moment matching.**

leads to a model size of $O(2k_{svd}n_p + m)k$ ). In practice, we have observed that a rank-one approximation is usually sufficient to provide a good accuracy. This model complexity is in contrast to that of a multi-point expansion approach, $O(c^{n_p}mk)$, in which $c$ samples per parameter are taken in the $n_p$ -dimensional variational parameter space.

# 5. Numerical Results

## 5.1 An RC network

We consider an RC network of 767 circuit unknowns subjected to two independent variational sources. We randomly vary the RC values of the circuit, and then extract the sensitivity matrices w.r.t. these two variational sources correspondingly. Finally, we applied the proposed algorithm to construct a variational reduced order model of size 37 (number of states). As outlined in Fig. 2, this model (approximately) matches up to 4$^{th}$ order multi-parameter moments w.r.t all parameters of the full model. For comparison, by taking 8 samples in the variational parameter space, we compute a 40-state multi-point expansion based model that matches up to 4$^{th}$ order moments w.r.t $s$ at each sample point.

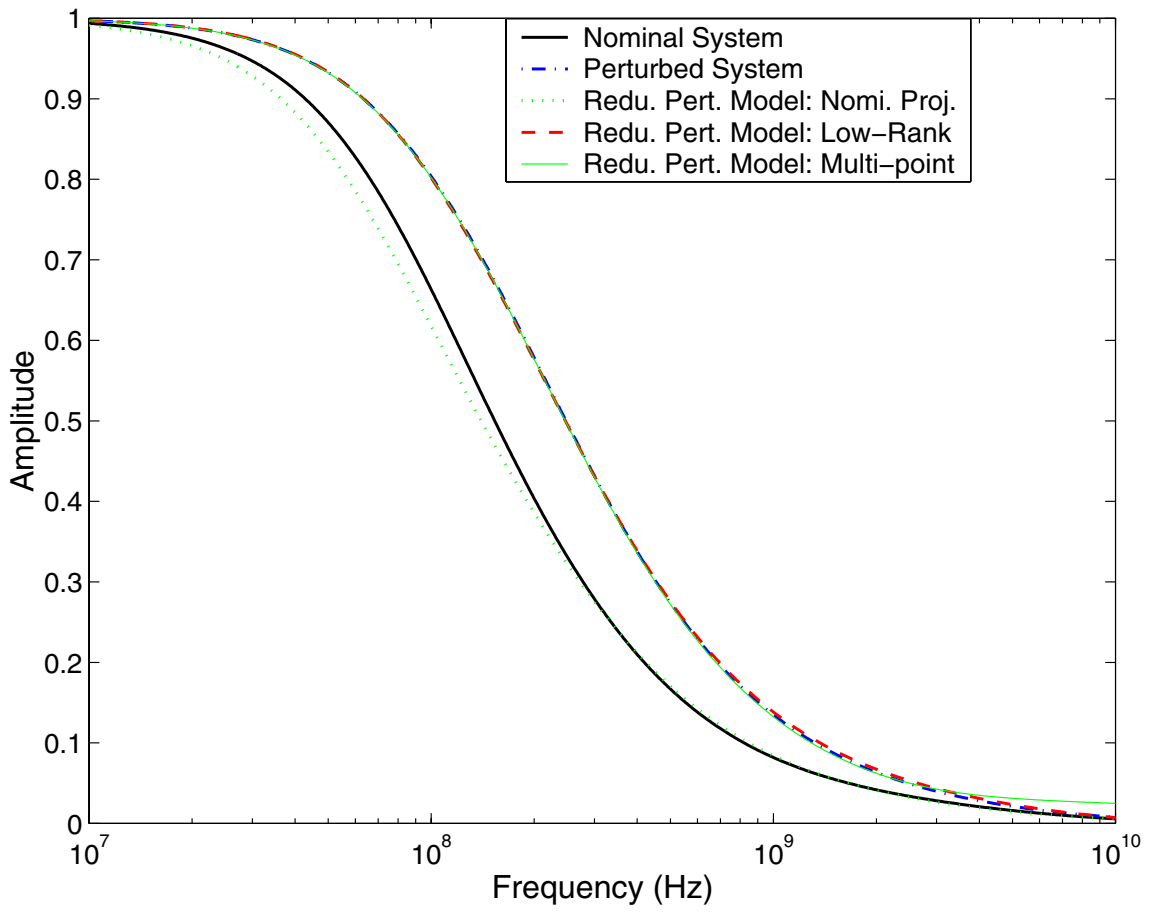

**Fig 3. A RC net for which the RC elements are subjected to a maximum 80% parametric variation.**

We evaluate the model accuracy for a perturbed network obtained by injecting up to 70% parametric variations into the nominal system. In Fig. 3, the transfer function from the voltage input to an observation node are plotted for five models: the nominal system, the perturbed system and three reduced models of the perturbed system. In addition to the aforementioned two reduced order models, we also include the third reduced order model in Fig. 3., which is obtained by projecting the variational model using a projection matrix obtained by applying PRIMA algorithm to the nominal system (matching 8 moments of $s$). It can be clearly seen from Fig. 3 that while the model based on the nominal projection matrix fails to capture the variation in transfer function, the other two reduced models become almost not indistinguishable with the perturbed full model in the plot.

## 5.2 A 4-port RLC network

Next, a two-bit bus is modeled as a coupled 4-port RLC network, where each line consists of 180 RLC segments. The size of MNA formulation for the bus is 1086. We follow the same procedure as in the previous example to generate parametric variations for two independent variational sources. Again, we consider three variational reduced order models. The first one has a size of 52, is obtained by employing the nominal projection matrix. The second model has a size of 144 and is computed by applying the proposed approach to (approximately) match the moments of all parameters (including cross-terms) up to 12$^{th}$ order. Among the moments matched, 52 are multi-port moments w.r.t $s$. The last reduced model is generated by taking three samples in the variational parameter space and applying the multi-point expansion. This model has a size of 156 and matches 52 multi-port moments w.r.t. $s$ at each sample point.

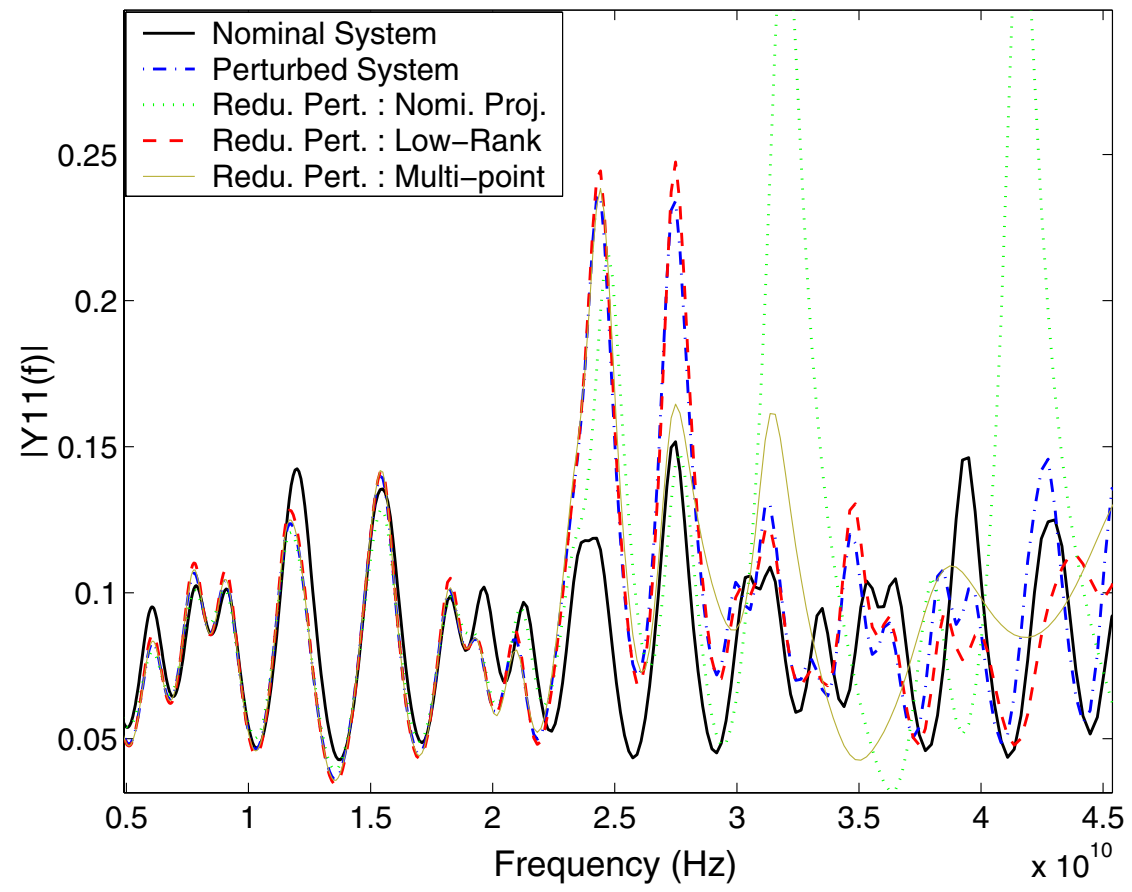

**Fig 4. A 4-port coupled RLC network. The perturbed system is subjected to a maximum 30% parametric variation.**

Various models are examined by evaluating a perturbed system subjected to a maximum of 30% parametric variation in Fig. 4. In the figure, we plot port admittance $Y_{11}$ based on the nominal full model, perturbed full model and three reduced order models. As shown in the plot, , the frequency domain response of a RLC circuit becomes more sensitive to parametric variations. It is also evident that for this case that building a variational model based upon only the information of the nominal system is far from adequate. However, by incorporating the variational information effectively into the model order reduction, the low-rank approximation based model is able to capture the variations in circuit response fairly accurately. In comparison, the larger multi-point-method model is not as accurate and the domain computational cost is three times larger.

## 5.3 Clock Trees

For more realistic examples, we consider two industrial RC networks, RCNetA and RCNetB. They are portions of a clock tree, and routed on three metal layers: M5, M6 and M7. RCNetA has 78 nodes while RCNetB 333. We consider three independent metal line width variations on these metal layers. In this case, the sensitivity matrices w.r.t metal line width variations are obtained by performing multiple parasitic extractions.

For RCNetA, we applied the low-rank based parametric model reduction to compute a reduced order model of size 29 while matching the moments of $s$ to the $4^{th}$ order and the rest of multi-parameter moments to the $2^{nd}$ order. We independently vary the three metal line widths up to 30% ($3\sigma$ variations) of the nominal values according to the normal distribution. For each perturbed circuit instance, we compare the reduced order model with the full model in terms of the 5 most dominant poles. The error distribution in these poles across all the instances is plotted in Fig. 5 (left plot). For this relatively small example, the error in these dominant poles is completely negligible. In Fig. 5 (right plot), we show the error in the most dominant pole as a function of M5 and M6 metal line widths (within -30% to 30% of their nominal values). This again confirms the accuracy of the reduced order model.

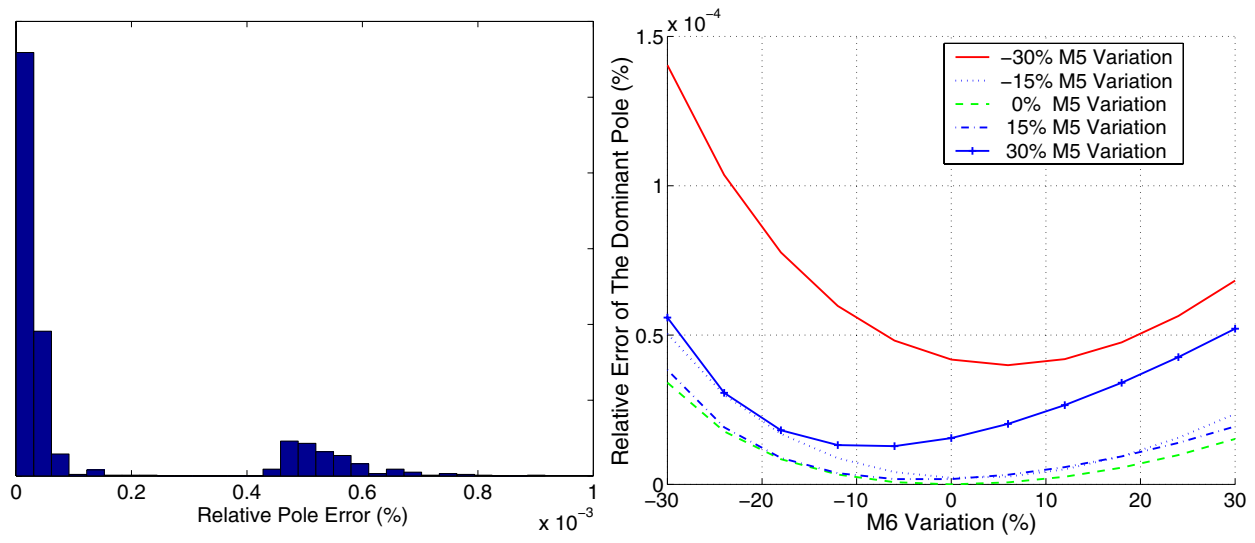

**Fig 5. Model accuracy for RCNetA. Left: error distribution in the 5 most dominant poles; Right: the error in the most dominant pole as a function of metal width variations.**

The same steps are taken to examine the model accuracy for RCNetB. We first compute a reduced order parametric model of size 40 while matching all the multi-parameter moments to the $3^{rd}$ order. Then, we independently vary the three metal line widths up to 30% ($3\sigma$ variations) of the nominal values using the normal distribution and plot the error in the 5 most dominant poles across all the instances in Fig. 6 (left plot). The maximum error out of 1000 poles is less than 0.12%. Next, we examine how the error in the most dominant pole changes as we vary the metal line widths of M5 and M6 in between -30% and 30%. As shown in Fig. 6 (right plot), the reduced order model is very accurate as the largest error is less than 0.3%.

## 6. Conclusion

We have shown that multi-point expansion in variational parameter space can be exploited to produce more compact parametric interconnect models than its single-point counterpart. Furthermore, we show that by combining low-rank matrix approximation and multi-parameter moment matching even more efficient parametric models can be constructed. Additionally, these variation-aware models are obtained at an almost linear cost and preserve the passivity.

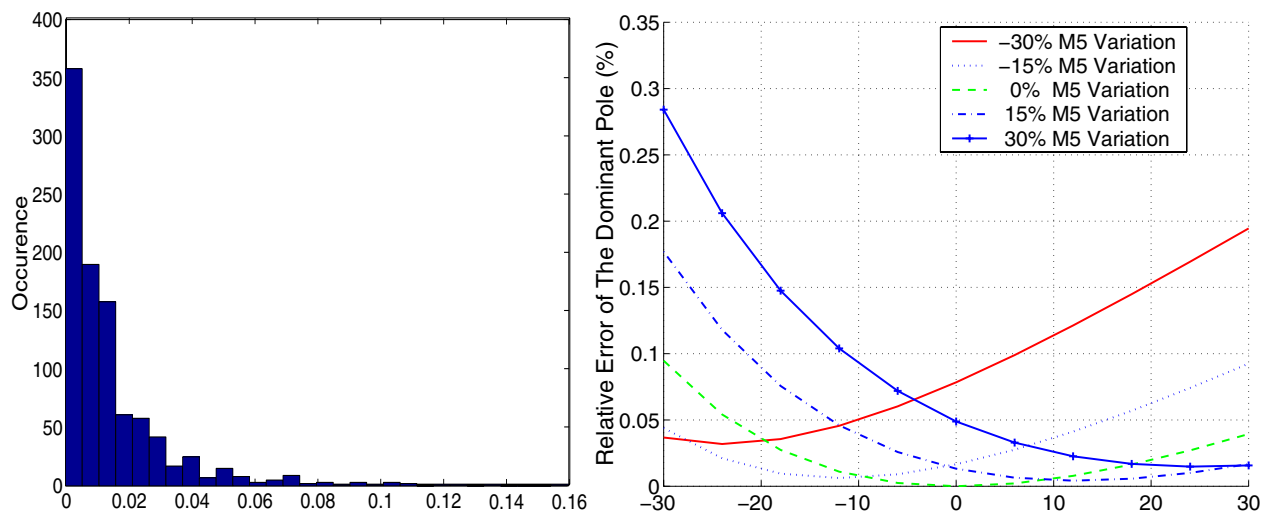

**Fig 6. Model accuracy for RCNetB. Left: error distribution in the 5 most dominant poles; Right: the error in the most dominant pole as a function of metal width variations.**